\documentclass[journal, twoside, onecolumn, draftclsnofoot,12pt]{IEEEtran}

\ifCLASSINFOpdf
  \usepackage[pdftex]{graphicx}

\else

\fi

\usepackage{multirow}

\begin{document}

\title{Group Communication Over LTE : A Radio Access Perspective}

\author{Juyeop~Kim,~\IEEEmembership{Member,~IEEE,}
        Sang~Won~Choi,~\IEEEmembership{Member,~IEEE,}
        Won-Yong~Shin,~\IEEEmembership{Member,~IEEE,}        
        Yong-Soo~Song,~\IEEEmembership{Member,~IEEE,}                        
        and~Yong-Kyu~Kim,~\IEEEmembership{Member,~IEEE}

\thanks{J. Kim, S.W. Choi, Y.-S. Song and Y.-K Kim are with the ICT Convergence Research Team, Korea Railroad Research Institute, Republic of Korea, email: \{jykim00, swchoi, adair, ygkim1\}@krri.re.kr, and W.-S. Shin is with the Department of Computer Science and Engineering, Dankook University, Republic of Korea, email: wyshin@dankook.ac.kr. S.W. Choi is the corresponding author of this article.}
}

\markboth{To appear in IEEE Communications Magazine}
{Submitted to IEEE Communications Magazine}

\maketitle

\begin{abstract}
Long Term Evolution (LTE), which has its root on commercial mobile communications, recently becomes an influential solution to future public safety communications. To verify the feasibility of LTE for public safety, it is essential to investigate whether an LTE system optimized for one-to-one communications is capable of providing group communication, which is one of the most important service concepts in public safety. In general, a number of first responders for public safety need to form a group for communicating with each other or sharing the common data for collaboration on their mission. In this article, we analyze how the current LTE system can support group communication in a radio access aspect. Based on the requirements for group communication, we validate whether each LTE-enabled radio access method can efficiently support group communication. In addition, we propose a new multicast transmission scheme, termed {\it index-coded Hybrid Automatic Retransmission reQuest (HARQ)}. By applying the index coding concept to HARQ operations, we show that the LTE system can provide group communication more sophisticatedly in terms of radio resource efficiency and scalability. We finally evaluate the performance of LTE-enabled group communication using several radio access methods and show how the proposed transmission scheme brings the performance enhancement via system level simulations.

\end{abstract}


\IEEEpeerreviewmaketitle

\section{Introduction}

\IEEEPARstart{M}{any} operators of commercial mobile communications nowadays provide personal data services along with a Long Term Evolution (LTE) system. Under this circumstance, many operators in other fields such as railway and public safety have begun to take into account the LTE system for special-purpose data communications, which is dedicated to accomplish a specific task in the specific fields. Many railway researches, including the Future Railway Mobile Communication System (FRMCS) project triggered by the International Union of Railway (UIC), estimate that LTE can meet the needs for transferring railway data in the long term \cite{LTER1}. Governments in many countries including the United States and the Republic of Korea have also been surveying how to utilize the LTE system for public safety communications \cite{PSLTE1}, \cite{PSLTE2}. In particular, the South Korean government has recently opened a request for the proposal of a demo business in July 2015 so that the public safety communications system based on the LTE is deployed. The main motive of this trend comes from that LTE network devices and terminals are ubiquitous and continuously upgraded according to the demand from vitalized commercial markets. From these facts, the operators can reduce a burden of both Operational Expenditure (OPEX) and Capital Expenditure (CAPEX) for fulfilling their needs for data communications.

To utilize the LTE system for special-purpose data communications, it is essential to investigate whether it is capable of providing group communication. Group communication is to disseminate the common voice or data context to multiple terminal users in a group and is a common form of the special-purpose data services. In many special-purpose scenarios, multiple staffs aim to accomplish a common mission and want to share various related information for collaboration. The representative application in the form of group communication is Push-To-Talk (PTT), where a user in a group sends a talk burst to the other listening users in half-duplex mode. Police officers or fire fighters form a group for each mission and communicate with each other through PTT for commanding and reporting. Locomotive engineers of a train, maintenance staffs on the track side, and station staffs also share the operational status of the train and negotiate train operations through PTT \cite{PSNetwork}.

For this reason, many researchers and engineers have recently discussed the requirements of a mobile communications system for supporting the group communication feature. According to the conclusion in the 3rd Generation Partnership Project (3GPP), the key performance measures for group communication are {\it latency} and {\it scalability} \cite{TS22468}, \cite{TS22179}. In a latency perspective, the setup time for a group call and the end-to-end delay of a group data dissemination are required to be within an allowable range regardless of the group size, so that every user in a group can experience a qualified group service. Based on the criteria in the requirement of TErrestrial Trunked RAdio (TETRA) mission critical voice systems, it is recommended to take less than 300ms from the moment that a user requests to join a group to the other moment that the user receives the first packet of the group data. It is also recommended that the end-to-end data transfer should be terminated within 150ms. In a scalability perspective, it is recommended to support such a case that the number of users in a group is unlimited, because massive staffs may belong to one group under public safety scenarios. In practice, a total of at least 2,000 users can participate and at most 500 users can be included in the same group \cite{TS22468}.

It is obvious that the typical way for transmitting data over LTE within the framework of unicast has a limitation to satisfy the above requirements. This motivates us to introduce a new system architecture and advanced data transmission schemes suitable for group communication. In a specification perspective, the 3GPP has recently been handling various specification items for supporting the group communication feature in LTE. Especially due to the needs of several governments for public safety communications, most of the technical specification groups in the 3GPP have focused on the group communication items in release 12 and 13 specifications, which include Group Communication System Enabler (GCSE), Single Cell Point-to-Multipoint (SC-PTM), and Mission Critical Push-To-Talk (MCPTT). In a research perspective, various studies have been conducted in the literature for providing efficient data communications to a group of terminals efficiently in mobile communications systems \cite{paperwork1}--\cite{paperwork4}. Most of their research are with respect to machine-to-machine communications, where massive machine nodes exist in a cell coverage and communicate with a base station.

The aim of this article is to introduce strong candidate methods for LTE-enabled group communication in a radio aspect. In the perspectives of scalability and latency, we provide an analysis of how the LTE-enabled radio access methods will perform in a group communication scenario. Furthermore, we propose a new multicast transmission scheme, which contributes to accomplish group communication in an efficient way. To design our scheme, we modify the Hybrid Automatic Retransmission reQuest (HARQ) operation based on a recent concept studied in the field of information theory, termed {\it index coding}. Through numerical evaluation, we validate that the proposed index-coded HARQ scheme can further enhance the performance of the LTE-enabled group communication.

\section{Radio Access Methods for LTE-enabled Group Communication}
We start by investigating how group communication can be realized through the air interface in the current LTE system. The most critical issue in the radio access level is how the common group data is disseminated to the wireless section between an eNodeB and a group of User Equipments (UEs) while fulfilling the requirements of {\it latency} and {\it scalability}. To understand the behavior of radio access in LTE, it is worth examining the structure and related procedures of LTE physical channels. Detailed descriptions of the physical channels are provided in release 12 and 13 specifications.

\begin{figure}[ht] 
\centering
\includegraphics[width=15.0cm]{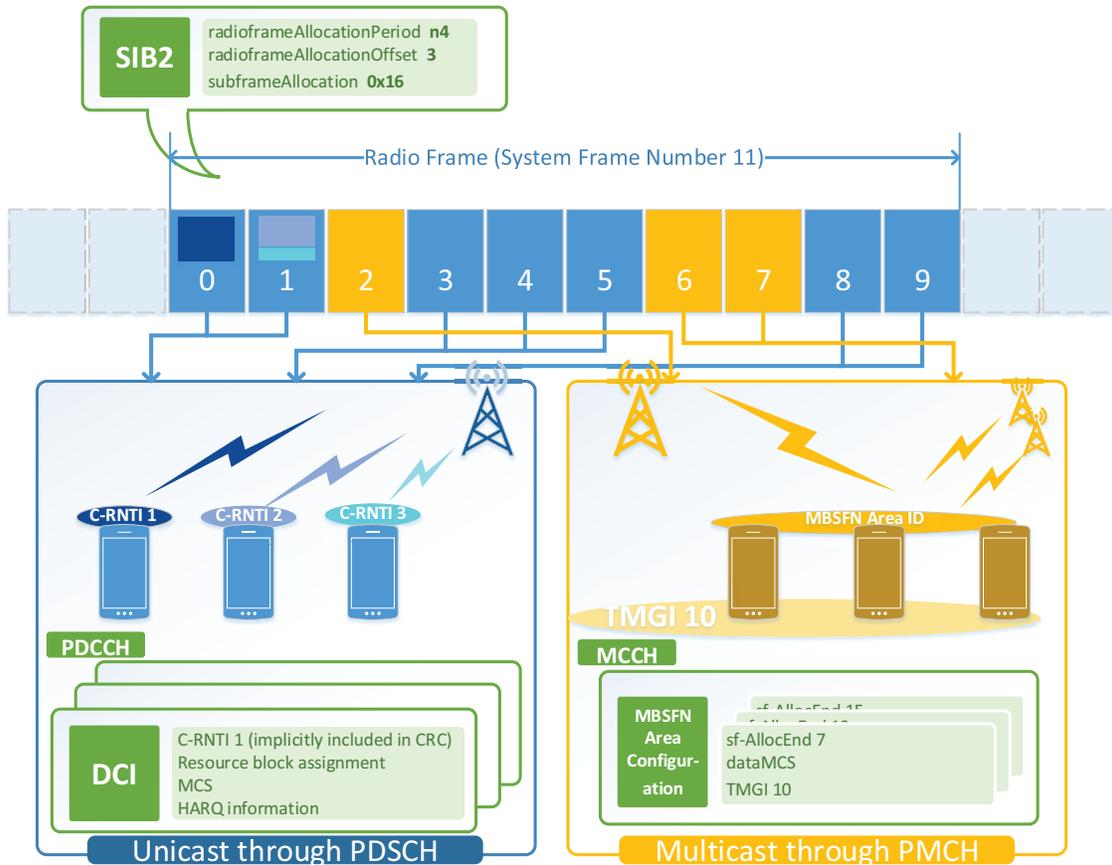}
\caption{An example of group communication through the PDSCH and PMCH.}
\label{pdschEmbms}
\end{figure}

\subsection{Release 12: Utilizing Conventional LTE Physical Channels}
The core specification item of release 12 in terms of group communication is GCSE. The conventional physical channels in LTE can be good mediums for providing group communication in some basic scenarios \cite{TS23468}. GCSE defines the requirements for group communication and proposes a system architecture on top of the existing physical channels. The LTE system of release 12 has two fundamental physical channels for transferring the data; Physical Downlink Shared Channel (PDSCH), which is commonly used for the normal unicast data, and Physical Multicast Channel (PMCH), which is designed for evolved Multimedia Broadcasting and Multicasting Service (eMBMS). Fig. \ref{pdschEmbms} describes an example of the frame structure in which the two physical channels coexist. In a certain radio frame consisting of 10 subframes, the two physical channels are switched on a basis of the subframe boundary. Based on an operational rule, a radio access network decides both portion and position of the subframes for the two physical channels and broadcasts the related control information to UEs through System Information Block Type 2 (SIB2). It is worth noting that the two physical channels are multiplexed only across the time domain, but not across the frequency domain. 

In a PDSCH subframe, each data is transferred only to a specific UE. Each UE communicating with its eNodeB has a Cell Radio Network Temporary Identifier (C-RNTI), which is unique in its belonging cell. During the physical layer encoding, each data is scrambled based on the C-RNTI of the receiving UE so that the UE can only decode the data successfully. In fact, a UE cannot usually notice whether the data for the other UEs passes through the PDSCH. A region of the orthogonal frequency division multiple access (OFDMA) radio resource allocated for the specific data in the PDSCH is indicated by a Downlink Control Information (DCI), which is transmitted through the Physical Downlink Control Channel (PDCCH). For every subframe, the UE initially attempts to decode the data in the PDCCH, and if it succeeds to decode its own DCI, then it performs decoding the data in the PDSCH based on the DCI. Importantly, the UE cannot decode the DCI for the other UEs. This is because the cyclic redundancy check (CRC) part of the DCI is scrambled based on the C-RNTI of the receiving UE and those who do not own the C-RNTI will suffer from the CRC check failure for the DCI. Thus, for group communication through the PDSCH, the group data should be duplicately transmitted for each UE in a group. As shown in Fig. \ref{pdschEmbms}, each of the three UEs receives the group data from the separate radio resource region within subframes {\it 0} and {\it 1}.

The advantage of group communication through the PDSCH is that the system can apply advanced link adaptation schemes utilized in LTE, such as adaptive modulation and coding (AMC), HARQ, and various multiple input multiple output (MIMO) schemes. Applying those technologies will definitely lead to a high spectral efficiency and an improved scalability. In addition, the end-to-end delay of the group data will be rather short when the PDSCH is used since the group data goes through the System Architecture Evolution (SAE) core network, which has a flat and all-IP architecture and is optimized in terms of minimizing latency. However, group communication through the PDSCH has a fundamental and critical problem that the group data should be duplicated as many as the number of UEs per group. The eNodeB should then allocate the radio resource separately for each UE in a group. This can be a bottleneck to satisfy the requirements of scalability and latency simultaneously, since the radio resource shortage and additional queuing delay at the eNodeB side may be beyond a certain critical level when there are many UEs in a group.

On the other hand, PMCH is optimized to broadcast the common data to multiple UEs. In the eMBMS system, the group data is carried through an eMBMS session identified by a Temporary Mobile Group Identity (TMGI) and is initially forwarded to an MBMS Coordination Entity (MCE). The MCE then multicasts the data to multiple eNodeBs which are grouped as a service area and configured to serve the eMBMS session. In a PMCH subframe, the eNodeBs simultaneously transmit the same physically encoded signals according to the scheduling by the MCE. At the same time, UEs interested in the eMBMS session attempt to combine the signals from the eNodeBs to decode the group data. Unlike the PDSCH, any UE willing to access the eMBMS session can receive the group data in the PMCH, because the data can be easily decoded based on the broadcasted information. The group data in the PMCH is scrambled with a Multimedia Broadcast Single Frequency Network (MBSFN) area ID that can be found in SIB13. An MBSFN area configuration message carrying the scheduling information for all the eMBMS sessions served in the cell is also available from the Multicast Control Channel (MCCH), which can be easily decoded with the information in SIB13. This allows the eNodeBs to disseminate the group data to multiple UEs with a single radio resource allocation.

It is advantageous that the amount of the radio resource consumed for group communication through the PMCH is independent of the number of UEs per group. No matter how many UEs exist in a group, the group data can be transferred to the UEs with a certain amount of radio resource through the PMCH. Group communication through the PMCH also enables the UEs in the cell edge region to achieve an improved performance via signal combining. However, the PMCH has a limitation to make a synergy effect along with various link adaptation schemes due to the lack of the uplink feedback channel. The eNodeB is then forced to utilize a robust Modulation and Coding Scheme (MCS) for the PMCH transmission, which results in a degraded spectral efficiency and gives a bad influence on the scalability. In addition, the granularity of the radio resource allocation in the PMCH is rather huge and thus is not suitable for multiplexing the small size data such as voice packets. Only the data from eMBMS sessions is allowed to be multiplexed in a PMCH subframe. Thus, there is no way to utilize the rest of the radio resource in the PMCH subframe when the amount of the group data to be sent is instantaneously small.\footnote{It is hard to change the portion of PDSCH and PMCH subframes in a short term, which requires to modify SIB2 frequently, thereby resulting in a burden to both UE and eNodeB sides.}

In a latency perspective, group communication through the PMCH performs properly while mostly satisfying the requirements as in the PDSCH case. In practice, the setup time for an eMBMS data bearer is generally similar to that for a normal unicast data bearer in the commercial mobile communications system. However, group communication through the PMCH may cause an additional queueing delay at the MCE/eNodeB sides when the subframe scheduled for transmitting the specific eMBMS session is far away from the current moment. The context of the MCCH cannot be modified during the MCCH modification period, referred to as SIB13. Thus, the scheduling for the PMCH cannot be changed during the MCCH modification period. This implies that the queueing time of group data will reach up to the MCCH modification period for the worst case.

\subsection{Release 13: Single Cell Point-to-Multipoint (SC-PTM)}
As reviewed above, both the physical channels in release 12 have their own problems with satisfying the requirements for group communication. To provide a fundamental solution for group communication, the 3GPP has recently started a specification item, so-called {\it SC-PTM} in release 13 \cite{SCPTM}. The SC-PTM is a new type of a radio access method dedicated to multicast through the PDSCH in a single cell. It can be regarded as a fusion of PDSCH and eMBMS. For an SC-PTM transmission, UEs in a group receive the group data through a common radio resource region in the PDSCH. This concept naturally allows the group data to be multiplexed with the normal unicast data within a PDSCH subframe and thus does not cause the problem of the radio resource granularity. 

\begin{figure}[ht] 
\centering
\includegraphics[width=15.0cm]{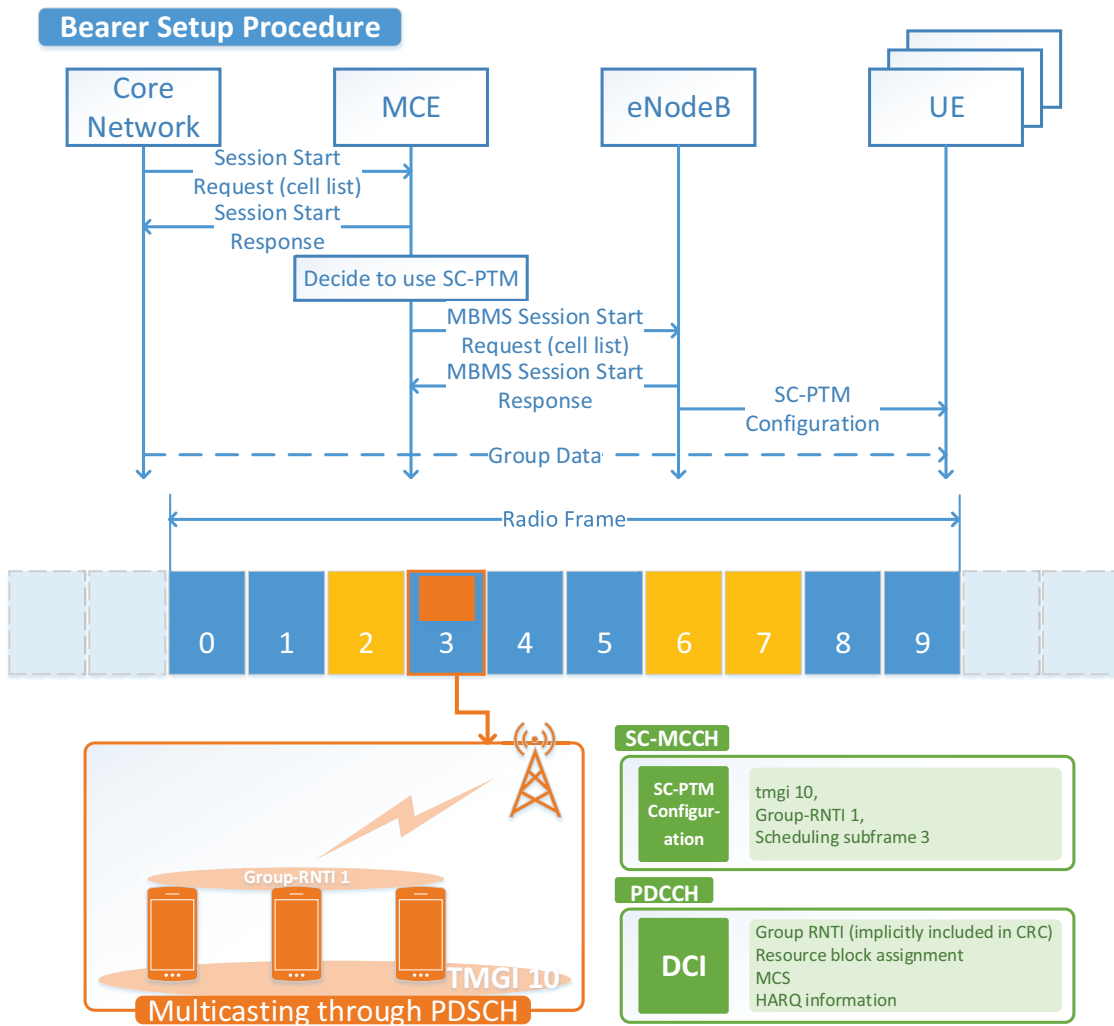}
\caption{The concept of SC-PTM.}
\label{SCPTM}
\end{figure}

Fig. \ref{SCPTM} depicts the details of the SC-PTM transmission. Instead of the C-RNTI, the SC-PTM transmission utilizes a common RNTI, so-called {\it group RNTI}, which is allocated to each TMGI. According to the cell list given from the core network, the MCE disseminates the group data to the corresponding eNodeBs. Each eNodeB then transmits the group data through the PDSCH based on its own scheduling and sends the corresponding DCI through the PDCCH simultaneously with the group RNTI. UEs can decode both the DCI and the group data successfully based on the pre-acquired group RNTI. The UEs in a group can acquire their group RNTI from an SC-PTM configuration message, which is periodically broadcasted through the Single Cell-MCCH (SC-MCCH) and provides the mapping between TMGIs and group RNTIs. Since the SC-PTM allows any UE to receive the group data as in the PMCH case, it only requires a single radio resource allocation for disseminating the group data without duplicated data transmissions.

The eNodeBs can utilize various link adaptation schemes for the SC-PTM because the uplink feedback channel corresponding to the PDSCH is available. Performance evaluation in \cite{SCPTM} showed that the SC-PTM transmission outperforms the transmission through the PMCH in terms of spectral efficiency. Although defining the uplink feedback channel for the SC-PTM is out of scope in release 13 specification, the numerical results in \cite{SCPTM} reveal that the uplink feedback channel still has a potential to bring a significant performance gain over the SC-PTM. Since the SC-PTM does not perform duplicated data transmissions for group communication, it consumes less amount of the radio resource even when there are relatively many UEs per group. In addition, the SC-PTM is attractive in an operator perspective, because it enables to manage the radio frame more flexibly by multiplexing the group data with the normal unicast data in any PDSCH subframe. 

In a latency point of view, group communication through the SC-PTM takes a smaller delay since the significant queueing delay, which may take place for both the PDSCH and PMCH cases, will not occur. Since the duplicated transmission is not needed for the SC-PTM case, the data queueing due to the radio resource shortage will hardly occur even when there are sufficiently many UEs per group. The group data is also transmitted through the PDSCH, in which the eNodeB performs scheduling per a subframe, and thus the queueing delay due to the scheduling of the MCE will not take place.


\begin{figure}[ht] 
\centering
\includegraphics[width=17.0cm]{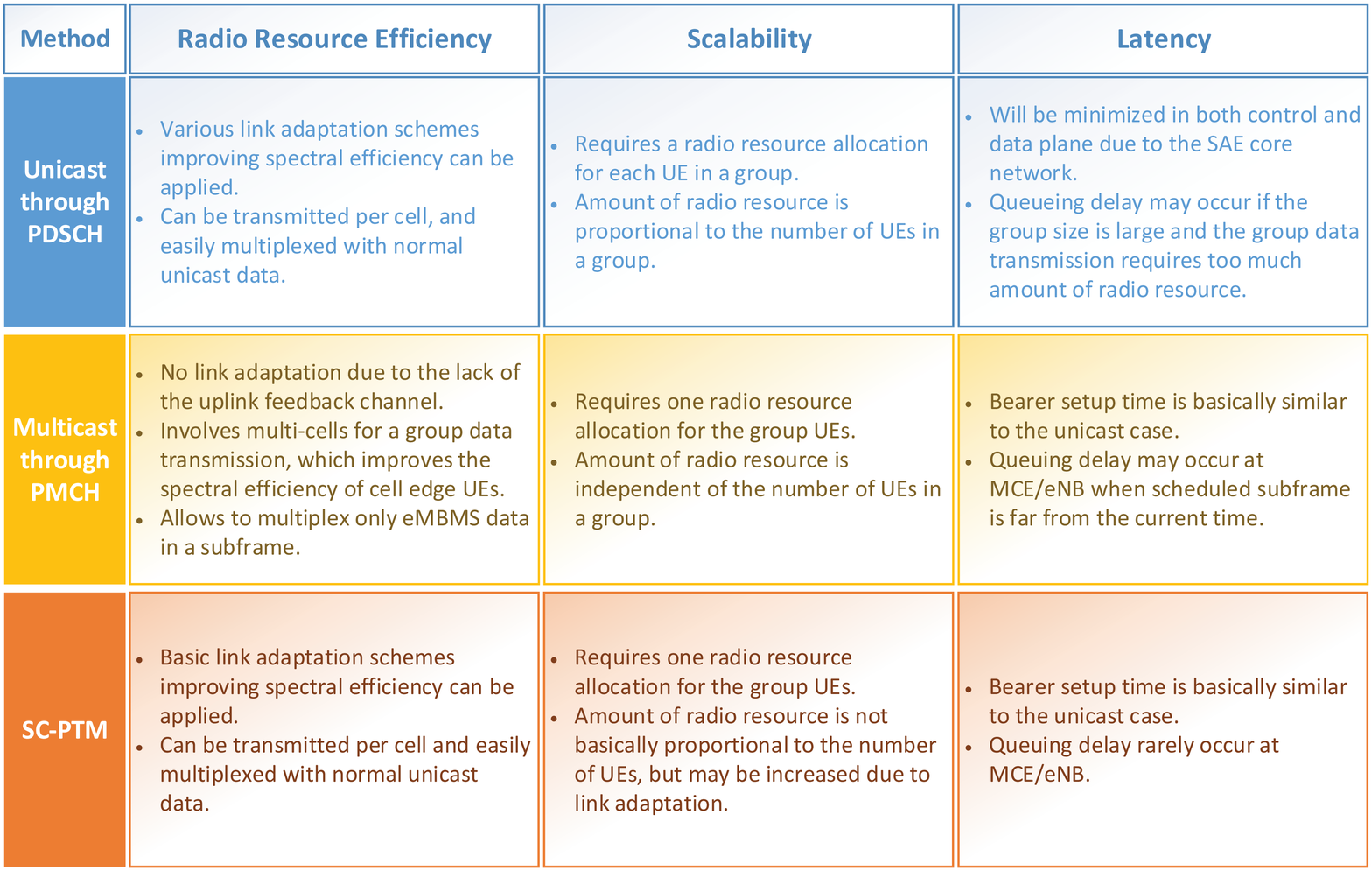}
\caption{The characteristics of various methods for group communication.}
\label{comparison}
\end{figure}

Fig. \ref{comparison} summarizes the characteristics of radio access methods. From the comparison, it is seen that the SC-PTM is a compromised solution between the existing physical channels. Unicast through the PDSCH and multicast through the PMCH are optimized to serve small and very large groups, respectively, while the SC-PTM is dominant for the mid-size groups. Thus, it is essential to utilize the SC-PTM to fulfill the requirements for group communication in every use scenario.

\section{New Paradigm: Index-coded HARQ for SC-PTM}

HARQ is one of the important link adaptation schemes, which plays a significant role of achieving a high spectral efficiency in the SC-PTM transmissions. The key HARQ operation is the physical-layer retransmission, which enables a transmitter to utilize the MCS with a higher data rate while suppressing the Block Error Rate (BLER) under 1\%. When HARQ is applied to the SC-PTM, its entity needs to perform retransmission according to multiple feedbacks from the UEs in a group. However, this HARQ retransmission can matter in terms of radio resource efficiency. Since the HARQ retransmission should take place when at least one NACK is sent from the UEs in the group, it may occur more frequently as the number of UEs per group increases. It would also be redundant for the UEs with ACK that the HARQ retransmission is performed through the SC-PTM, due to the inherent nature of this retransmission method.
\begin{figure}[ht] 
\centering
\includegraphics[width=17.0cm]{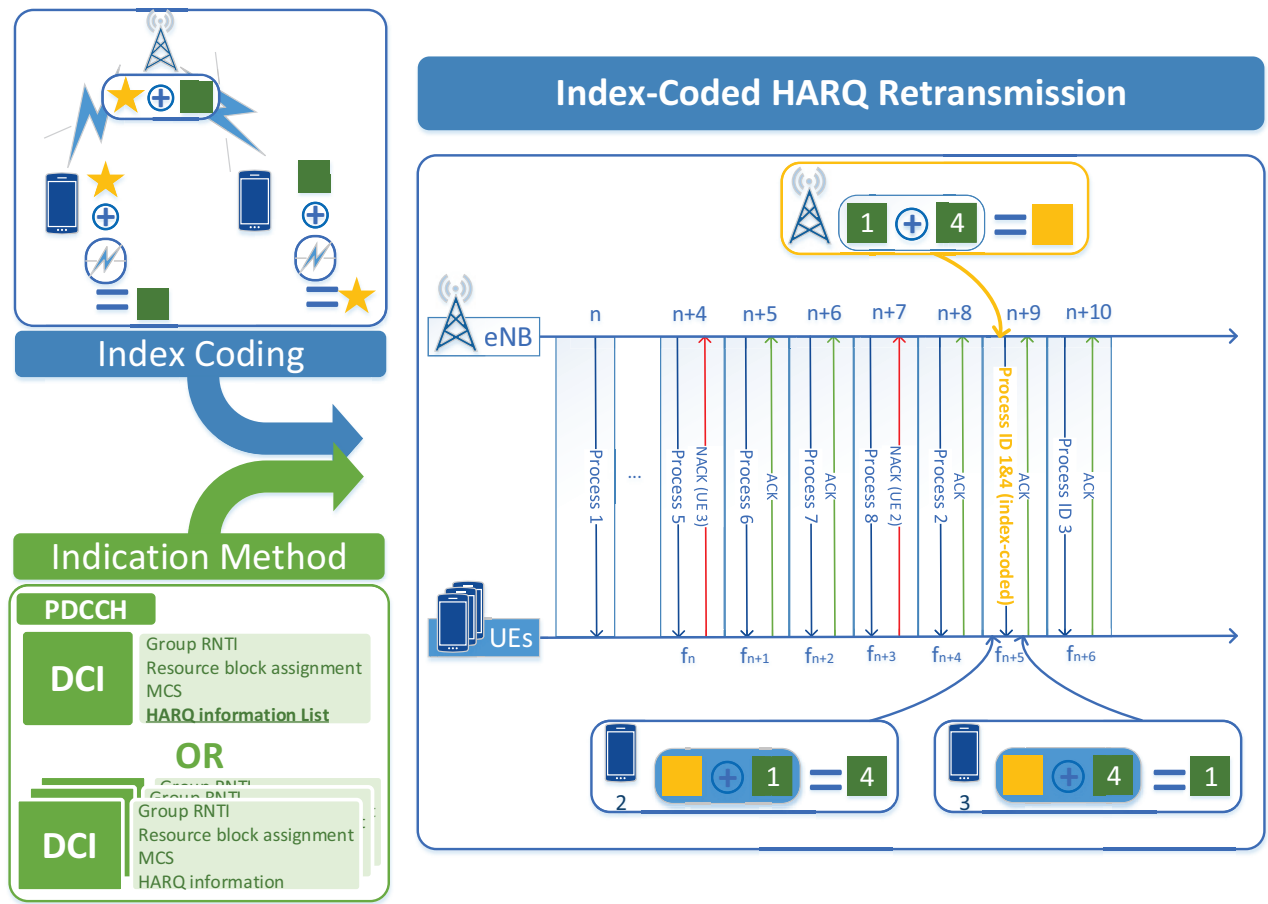}
\caption{An example of index-coded HARQ retransmissions for the SC-PTM.}
\label{indexCoding}
\end{figure}

To overcome this inefficiency, the issue of the retransmission to multiple receivers can be combined with a new coding scheme, termed {\it index coding}, studied in the field of information theory. The transmitter sends data through a noiseless broadcasting network to multiple receivers, each knowing some data a prior, referred to as {\it side information}. Then, one can exploit this side information to reduce the number of coded data to be sent by the transmitter, for all receivers to decode their requested data. This concept is known as the index coding problem and was originally introduced by Birk and Kol \cite{IC1}, motivated by a satellite broadcasting application. Interest in index coding has further been increased owing to two more recent developments \cite{IC2}, \cite{IC3}.

Based on the sophisticated philosophy of the index coding, we can enhance the HARQ retransmission for multiple UEs in the sense of diminishing the number of retransmissions. In LTE, the HARQ entity at the eNodeB side operates based on several HARQ processes in parallel, and each HARQ process is responsible to transfer a Transport Block (TB), which is the minimal unit of data upon HARQ operations. The eNodeB is aware of the group's reception status (ACK/NACK) for each HARQ process from the received HARQ feedbacks. Based on the ACK/NACK information, the eNodeB can select proper HARQ processes whose TBs can be index-coded for a retransmission. According to the principle of index coding, a UE can retrieve a TB from {\it m} index-coded TBs assuming that the UE already has the {\it m-1} TBs. In other words, each NACK UE can retrieve the TB that it wants to receive from the index-coded TBs only if there is no common NACK UE among the selected HARQ processes. Thus, the index coding enables simultaneous retransmissions for several HARQ processes by carefully selecting the HARQ processes whose sets of the NACK UEs are disjoint.

Fig. \ref{indexCoding} shows an example of the index-coded HARQ, where there are 3 UEs receiving the group data through the SC-PTM. In subframe {\it n+4}, UE {\it 3} sends the NACK for HARQ process {\it 1}. In subframe {\it n+7}, UE {\it 2} sends the NACK for HARQ process {\it 4}. After collecting a certain amount of HARQ feedbacks from the UEs, the eNodeB checks whether there is a possible combination of the HARQ processes for the index-coded retransmission. In this case, HARQ processes {\it 1} and {\it 4} can be index-coded, because UEs {\it 2} and {\it 3} have successfully decoded the TBs of HARQ processes {\it 1} and {\it 4}, respectively. The eNodeB combines the TBs of HARQ processes {\it 1} and {\it 4} by applying an exclusive OR (XOR) operation and transmits the index-coded TBs. After completing the receiving procedure in the PDSCH, UE {\it 2} applies an XOR operation to the decoded data with the TB of HARQ process {\it 1} and retrieves the TB of HARQ process {\it 4}. Similarly, UE {\it 3} retrieves the TB of HARQ process {\it 1} from the index-coded TBs. Consequently, the index-coded HARQ can conduct the two HARQ retransmissions with a single radio resource allocation, whereas two radio resource allocations are needed to perform the two HARQ retransmissions through the conventional HARQ operation with no index coding.

The key point of using the index coding to the SC-PTM is to reduce the amount of the radio resource consumed for HARQ retransmissions. More specifically, this leads to a reduced amount of the radio resource for disseminating the group data, thus resulting in an improved scalability for group communication. The index-coded HARQ also reduces the frequency of duplicated reception and disuse by UEs with ACK. In addition, applying the index-coded HARQ requires a minor change of the conventional LTE system in a protocol aspect by adding the information for index-coded TBs that can be naturally sent through the PDCCH. Specifically, either multiple sets of HARQ information can belong to a DCI or multiple DCIs can be sent to indicate which TBs are index-coded.

\section{Performance of group communication}
We evaluate the performance of group communication in a scalability aspect. Our aim is to show {\it i)} how much scalable each radio access method is for group communication and {\it ii)} how to improve the scalability using the proposed scheme. To evaluate the scalability of the PMCH, we conduct a numerical analysis based on the framework in \cite{SCPTM}. For both the PDSCH and SC-PTM, we evaluate the scalability in LTE-based simulation environments.

\subsection{System Assumptions}

For numerical evaluation, we basically use the LTE system along with the system parameters in \cite{SCPTM}, which are indeed commonly used for evaluating public safety scenarios. In addition, we assume that the voice traffic is used, since the voice PTT is the main application in public safety. The system parameters for simulation are summarized in detail in Table \ref{simParameter}. 

\begin{table*}
\caption{System parameters for simulation}
\begin{center}
\begin{tabular}{||c|p{5.0cm}||}
   \hline
   System parameters & Values \\
   \hline\hline
   Channel model & ITU, rural macro-cell \\
   \hline
   eNodeB layout & 19 hexagonal cells\\
   \hline
   Distance between eNodeBs & 1732m \\
   \hline
   Subcarrier spacing & 15kHz \\
   \hline
   Carrier frequency and bandwidth & 800MHz, 10MHz BW\\
   \hline
   UE speed & 3km/h\\
   \hline
   Duplex & FDD \\
   \hline
   eNodeB antenna gain & 15dBi \\
   \hline
   eNodeB output power & $46$dBm \\
   \hline
   UE distribution & Uniform drop in the cell coverage region \\
   \hline
   Traffic model & Voice \\
   \hline
   Downlink transmission scheme & TxD\\
   \hline
   Antenna configuration & 2x2\\
   \hline
  HARQ type & Chase combining up to 3 retransmissions \\
  \hline
  Rate adaptation & Feedback from group in the worst radio condition will be ignored, with 1\% BLER\\
  \hline
   Number of OFDM symbols reserved for PDCCH & 2\\
  \hline   
\end{tabular}
\end{center}
\label{simParameter}
\end{table*}

For the quantitative analysis of the scalability, we define a performance measure, called {\it group capacity}, which is the maximal number of groups that a cell can support with a given group size. The group capacity can be calculated as the total amount of the radio resource within the inter-arrival time of the group data traffic normalized by the average amount of the radio resource needed for disseminating a piece of group data to the group. Through simulations, we evaluate the amount of the radio resource that a HARQ process consumes for transmitting/retransmitting a TB while satisfying the 1\% BLER criterion.

\begin{figure}[ht] 
\centering
\includegraphics[width=17.0cm]{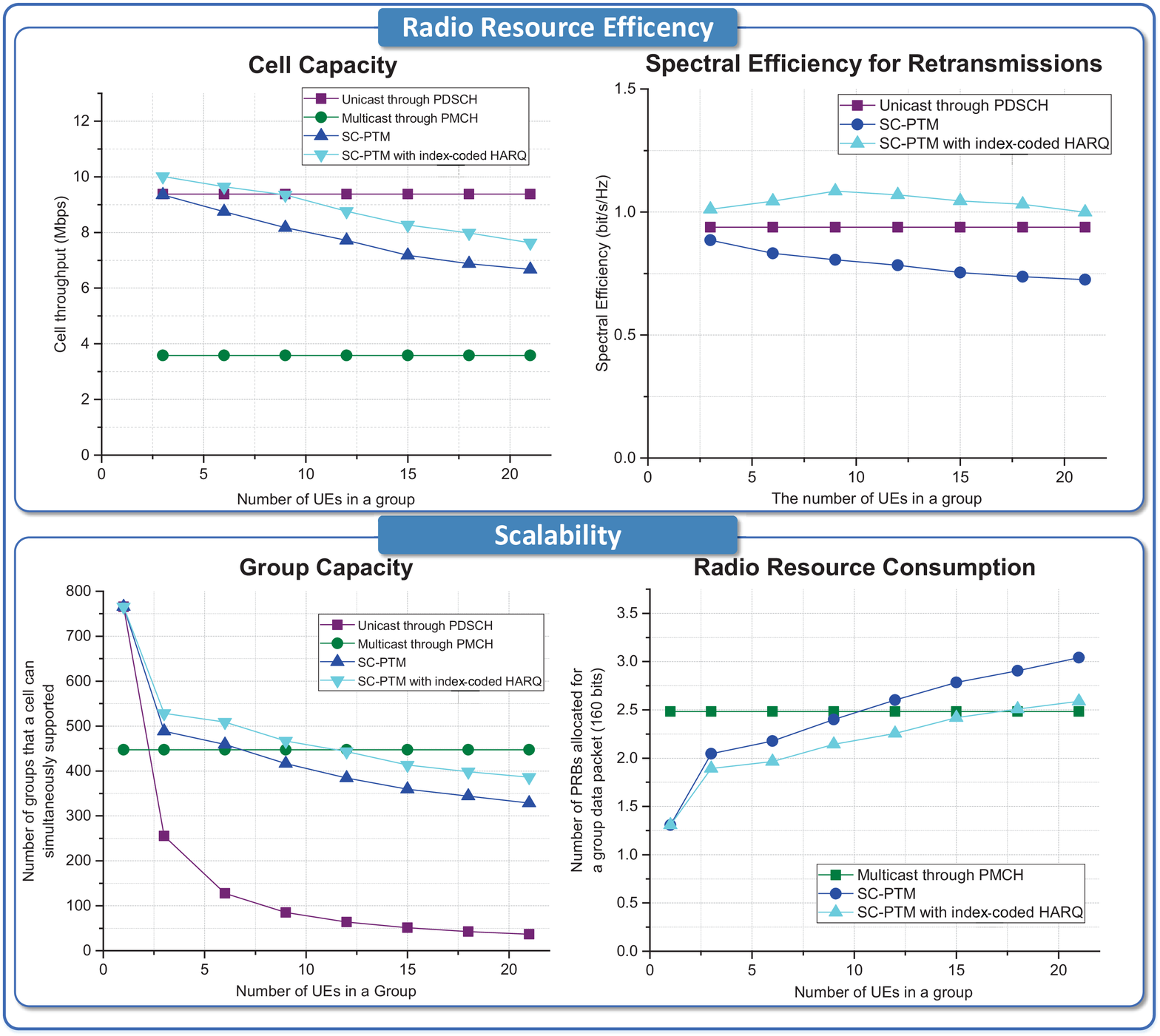}
\caption{Performance comparison of the various radio access methods.}
\label{performance}
\end{figure}

\subsection{Performance Evaluation}

%
%
%
Fig. \ref{performance} shows the performance of each radio access method in terms of scalability and radio resource efficiency. Unicast through the PDSCH outperforms the other methods for most cases with respect to the cell capacity. However, unicast through the PDSCH has the worst performance on the group capacity even when the group size is very small. This is because this unicast method wastes a substantial amount of radio resource for duplicated data transmissions. Multicast through the PMCH has the worst performance on the cell capacity, but has a higher group capacity than unicast through the PDSCH for most cases (even a higher group capacity than the SC-PTM when there are relatively many UEs per group). This implies that the best method in maximizing the group capacity is dependent on the group size. The LTE system will prefer to serve either a small group size through the SC-PTM or a large group size, which can be up to 500 users, through the PMCH. It is also worth noting that all the subframes are configured for the PMCH and there is no subframe configured for the PDSCH. Since the radio frame is generally composed of both PMCH and PDSCH subframes in practice, the group capacity for the multicast case through the PMCH will be lessened as much as the portion of the PMCH subframes. Therefore, a cross point of the group capacity curves between the PMCH and the SC-PTM will vary depending on the frame configuration.

Fig. \ref{performance} also shows how the SC-PTM with index-coded HARQ performs. It turns out that the group capacity can be improved up to 17.5\%  using the proposed scheme. The performance gain gets larger as the number of UEs per group increases. This is because the large group size leads to more NACKs, resulting in more frequent HARQ retransmissions, and thus index-coded transmission occurs more frequently. We remark that the index-coded HARQ compensates a vulnerable point of the SC-PTM, which enables to significantly extend the availability of the SC-PTM. It is shown that the SC-PTM without index-coded HARQ is applicable when there are fewer than 6 UEs per group, whereas the SC-PTM with index-coded HARQ can be applied for the case having 12 UEs per group. 

Moreover, the index-coded HARQ can improve the performance over the SC-PTM in terms of radio resource efficiency. The result indicates that the SC-PTM with index-coded HARQ consumes the radio resource for retransmission more efficiently than that of the SC-PTM without index coded HARQ, (even than the case of unicast through the PDSCH). In addition, the cell capacity of the SC-PTM, depending heavily on the spectral efficiency, is higher than that of unicast through the PDSCH when the group size is relatively small. This reveals that the enhancement of the group capacity by the index-coded HARQ comes mainly from the improved radio resource efficiency.

\section{Concluding Remarks}
It was comprehensively verified that group communication is one of the most important and widely used application for public safety, as one-to-one voice communication is for commercial mobile communications. As an applicable range of LTE is recently extended to various fields including public safety, it is essential to account for whether the LTE system can fulfill the requirements for group communication in an efficient way. In this circumstance, this article sheded light on the technical aspect of LTE in terms of providing group communication. In scalability and latency perspectives, the release 12 LTE system was shown to some extent to support group communication by using both unicast (PDSCH) and multicast (PMCH) channels. Furthermore, SC-PTM defined in release 13 was turned out to be the compromised solution of the existing physical channels and to fulfill the requirements for group communication more smoothly. The proposed index-coded HARQ led to a new paradigm of retransmission to multiple receivers, where it can further enhance the scalability of SC-PTM, and was validated via numerical evaluation.


\section*{Acknowledgment}
This work was supported by ICT R\&D program of MSIP/IITP. [B0101-15-1361, Development of PS-LTE System and Terminal for National Public Safety Service].

\ifCLASSOPTIONcaptionsoff
  \newpage
\fi

%

%
%
%




\newpage

\section*{Biographies}

\begin{IEEEbiographynophoto}{Juyeop Kim}
Juyeop Kim (jykim00@krri.re.kr) is an senior researcher in ICT Convergence Team at Korea Railroad Research Institute. He received his M.S. and Ph.D. in electrical engineering and computer science from KAIST in 2010. His current research interests are railway communications systems, group communications and mission critical communications.
\end{IEEEbiographynophoto}

\begin{IEEEbiographynophoto}{Sang Won Choi}
Sang Won Choi (swchoi@krri.re.kr) received his M.S. and Ph.D. in electrical engineering and computer science from 
KAIST in 2004 and 2010, respectively. He is currently a senior researcher in ICT Convergence Research Team. 
His research interests include mission critical communications, mobile communication, communication signal processing, and
multi-user information theory. He was the recipient of a Silver Prize at Samsung Humantech Paper Contest in 2010.
\end{IEEEbiographynophoto}

\begin{IEEEbiographynophoto}{Won-Yong~Shin}
Won-Yong Shin (S'02-M'08) received the B.S. degree in electrical engineering from Yonsei University, Seoul, Korea, in 2002. He received the M.S. and the Ph.D. degrees in electrical engineering and computer science from Korea Advanced Institute of Science and Technology (KAIST), Daejeon, Korea, in 2004 and 2008, respectively. From February 2008 to April 2008, he was a Visiting Scholar in the School of Engineering and Applied Sciences, Harvard University, Cambridge, MA. From September 2008 to April 2009, he was with the Brain Korea Institute and CHiPS at KAIST as a Postdoctoral Fellow. From August 2008 to April 2009, he was with the Lumicomm, Inc., Daejeon, Korea, as a Visiting Researcher. In May 2009, he joined Harvard University as a Postdoctoral Fellow and was promoted to a Research Associate in October 2011. Since March 2012, he has been with the Division of Mobile Systems Engineering, College of International Studies and the Department of Computer Science and Engineering, Dankook University, Yongin, Korea, where he is currently an Assistant Professor. His research interests are in the areas of information theory, communications, signal processing, mobile computing, big data analytics, and online social networks analysis.

Dr. Shin has served as an Associate Editor for the IEICE TRANSACTIONS ON FUNDAMENTALS OF ELECTRONICS, COMMUNICATIONS, COMPUTER SCIENCES, for the IEIE TRANSACTIONS ON SMART PROCESSING and COMPUTING, and for the JOURNAL OF KOREA INFORMATION AND COMMUNICATIONS SOCIETY. He also served as an Organizing Committee for the 2015 IEEE Information Theory Workshop.
\end{IEEEbiographynophoto}

\begin{IEEEbiographynophoto}{Yong-Soo Song}
Yong-soo Song(adair@krri.re.kr) received the Master´s degree in Electrical Engineering from Yonsei University in 2004. He has been with KRRI(Korea Railraod Research Institute) since 2004. He is working toward his Ph.D in Electrical Engineering from Yonsei University. His current research interests are in Cell planning and handover in LTE railway.
\end{IEEEbiographynophoto}

\begin{IEEEbiographynophoto}{Yong-Kyu Kim}
Yong-Kyu Kim (ygkim1@krri.re.kr) received his M.S. in Electronic engineering from Dankook University, Korea, in 1987 and his DEA and Ph.D. in automatic and digital signal processing from Institute National Polytechnique de Lorraine, France, in 1993 and 1997, respectively. He is currently a executive researcher in ICT convergence research team at Korea Railroad Research Institute. His research interests are in automatic train control, communication based train control, and driverless train operation. 
\end{IEEEbiographynophoto}


\begin{thebibliography}{1}

\bibitem{LTER1}
J. Kim, S.W. Choi, Y.-S. Song, Y.-K. Yoon, and Y. K. Kim, ``Automatic train control over LTE: Design and performance evaluation," {\it IEEE Communications Magazine}, vol. 53, no. 10, pp. 102--109, Oct. 2015. 


\bibitem{PSLTE1}
T. Doumi, M. F. Dolan, S. Tatesh, A. Casati, G. Tsirtsis, K. Anchan, and D. Flore, ``LTE for public safety networks," {\it IEEE Communications Magazine}, vol. 51, no. 2, pp. 106--112, Feb. 2013.

\bibitem{PSLTE2}
R. Ferrus, O. Sallent, G. Baldini, and L. Goratti, ``LTE: The technology driver for future public safety communications," {\it IEEE Communications Magazine}, vol. 51, no. 10, pp. 154--161, Oct. 2013.

\bibitem{PSNetwork}
K. Balachandran, K. C. Budka, T. P. Chu, T. L. Doumi and J. H. Kang, "Mobile responder communication networks for public safety," {\it IEEE Communications Magazine}, vol. 44, no. 1, pp. 56--64, Jan. 2006.

\bibitem{TS22468}
3GPP TS 22.468 v12.1.0, {\it Technical Specification Group Services and System Aspects; Group Communication System Enablers for LTE (GCSE\_LTE)}, 2014. 

\bibitem{TS22179}
3GPP TS 22.179 v13.2.0, {\it Technical Specification Group Services and System Aspects; Mission Critical Push To Talk (MCPTT) over LTE; Stage 1}, 2015. 

\bibitem{paperwork1}
T. Kwon and J. W. Choi, ``Multi-group random access resource allocation for M2M devices in multicell systems," {\it IEEE Communications Letters}, vol. 16, no. 6, pp. 834--837, June 2012.

\bibitem{paperwork2}
C. H. Wei, R. G. Cheng, and S. L. Tsao, ``Performance analysis of group paging for machine-type communications in LTE networks," {\it IEEE Transactions on Vehicular Technology}, vol. 62, no. 7, pp. 3371--3382, Sep. 2013.

\bibitem{paperwork3}
K. Zheng, F. Hu, W. Wang, W. Xiang, and M. Dohler, ``Radio resource allocation in LTE-advanced cellular networks with M2M communications," {\it IEEE Communications Magazine}, vol. 50, no. 7, pp. 184--192, July 2012.

\bibitem{paperwork4}
R. Sivaraj, A. K. Gopalakrishna, M. G. Chandra, and P. Balamuralidhar, ``QoS-enabled group communication in integrated VANET-LTE heterogeneous wireless networks," {\it 2011 IEEE 7th International Conference on Wireless Mobile Computing, Network and Communications}, pp. 17--24, Oct. 2011.

\bibitem{TS23468}
3GPP TS 23.468 v12.5.0, {\it Technical Specification Group Services and System Aspects; Group Communication System Enablers for LTE (GCSE\_LTE); Stage 2}, 2015. 

\bibitem{SCPTM}
3GPP TR 36.890 v13.0.0, {\it Technical Specification Group Radio Network; Evolved Universal Terrestrial Radio Access (E-UTRA); Study on single-cell point-to-multipoint transmission for E-UTRA}, 2015. 

\bibitem{IC1}
Y. Birk and T. Kol, ``Coding-on-demand by an informed source (ISCOD) for efficient broadcast of different supplemental data to caching clients," {\it IEEE Transactions on Information Theory}, vol. 52, no. 6, pp. 2825--2830, June 2006.

\bibitem{IC2}
Z. Bar-Yossef, Y. Birk, T. S. Jayram, and T. Kol, ``Index coding with side information," {\it IEEE Transactions on Information Theory}, vol. 57, no. 3, pp. 1479--1494, March 2011.

\bibitem{IC3}
S. El Rouayheb, A. Sprintson, and C. Georghiades, ``On the index coding problem and its relation to network coding and matroid theory," {\it IEEE Transactions on Information Theory}, vol. 56, no. 7, pp. 3187--3195, July 2010.

\end{thebibliography}
\end{document}